% Revised Version: October 30, 1996 Certain references have been 
%                  added.
%
%
% Martin Bucher and Yong Zhu, Non-Gaussian Isocurvature Perturbations 
% From Goldstone Modes Generated During Inflation
% October 28, 1996 Final Version
% Uses phyzzx.tex and epsf.tex macros
%
\input epsf.tex 
\input phyzzx 

\def\eatOne#1{}

\global\newcount\figno \global\figno=0

\def\ifundef#1{\expandafter\ifx%
\csname\expandafter\eatOne\string#1\endcsname\relax}

\def\mfig#1#2{\global\advance\figno by1%
\relax#1\the\figno%
\warnIfChanged#2{\the\figno}%
\edef#2{\the\figno}%
\reflabeL#2%
\ifWritingAuxFile\immediate\write\auxfile{\noexpand\xdef\noexpand#2{#2}}\fi%
}

% \LoadFigure is used to put a figure into the text.  Its first argument
% is the symbolic name for the figure (if it isn't defined, a new number
% will be assigned);  the second argument is a caption;
% the third argument size information in the form 
% \epsfxsize=3.0in\epsfysize=3.5in (this argument may be blank and
% may contain any valid preparatory argument used by the epsf package);
% the fourth and last argument is the name of the file which contains the 
% figure.
% The macro is basically just a front-end for \epsfbox; its purpose is 
% to allow figures to be switched from placement in the running text
% to placement on a separate page at the end of the text.  This choice
% is made using the flag \FiguresInText{true,false}; in the latter case,
% figures are placed at the end, size information is ignored (figures
% will be full-size), and the captions are listed separately on a page
% when the \listfigs command is invoked, followed by the figures, each
% on a separate page.  
%  The epsf package must be loaded by the user.
\newif\ifFiguresInText\FiguresInTexttrue
%\newif\if@FigureFileCreated 
\newwrite\capfile
\newwrite\figfile

\def\PlaceTextFigure#1#2#3#4{%
\vskip 0.5truein%
#3\hfil\epsfbox{#4}\hfil\break%
\hfil\vbox{Figure #1. #2}\hfil%
\vskip10pt}
\def\PlaceEndFigure#1#2{%
\epsfysize=\vsize\epsfbox{#2}\hfil\break\vfill\centerline{Figure #1.}\eject}

\def\LoadFigure#1#2#3#4{%
\ifundef#1{\phantom{\mfig{}#1}}\fi
\ifWritingAuxFile\immediate\write\auxfile{\noexpand\xdef\noexpand#1{#1}}\fi%
\ifFiguresInText% Figure is immediate
\PlaceTextFigure{#1}{#2}{#3}{#4}%
\else% Figure is at the end
\if@FigureFileCreated\else%
\immediate\openout\capfile=\jobname.caps%
\immediate\openout\figfile=\jobname.figs%
\fi%
\immediate\write\capfile{\noexpand\item{Figure \noexpand#1.\ }#2.}%
\immediate\write\figfile{\noexpand\PlaceEndFigure\noexpand#1{\noexpand#4}}%
\fi}

\def\listfigs{\ifFiguresInText\else%
\vfill\eject\immediate\closeout\capfile%\parindent=20pt
\immediate\closeout\figfile%
\centerline{{\bf Figures}}\bigskip\frenchspacing%
\input \jobname.caps\vfill\eject\nonfrenchspacing%
\input\jobname.figs\fi}

% An .aux file --- for forward references...
\newif\ifWritingAuxFile
\newwrite\auxfile
\def\SetUpAuxFile{%
\xdef\auxfileName{\jobname.aux}%
% Read it in if it exists
\inputAuxIfPresent{\auxfileName}%
% Now write a new one.
\WritingAuxFiletrue%
\immediate\openout\auxfile=\auxfileName}

% Some generally useful notation

% end of header.tex
\baselineskip=18pt
\parskip=0pt
\hoffset=0.2truein
\hsize=6truein
\voffset=0.1truein
\def\TITLEPAGE{\frontpagetrue}
\def\PUPT#1{\hbox to\hsize{\tenpoint \baselineskip=12pt
        \hfil\vtop{
        \hbox{\strut PUPT-96-#1}
}}}
\def\ITP#1{\hbox to\hsize{\tenpoint \baselineskip=12pt
        \hfil\vtop{
        \hbox{\strut ITP-SB-96-#1}
}}}
\def\PRINCETON{
\centerline{${}^1$ Department of Physics}
\centerline{Princeton University}
\centerline{Princeton, New Jersey 08544}}
\def\SB{
\centerline{${}^2$ Institute for Theoretical Physics}
\centerline{State University of New York}
\centerline{Stony Brook, New York 11794}}

\def\TITLE#1{\vskip .0in \centerline{\fourteenpoint #1}}
\def\AUTHOR#1{\vskip .1in \centerline{#1}}

\def\ABSTRACT#1{\vskip .1in \vfil \centerline{\twelvepoint
\bf Abstract}
   #1 \vfil}
\def\ENDTITLEPAGE{\vfil\eject\pageno=1}
\hfuzz=5pt
\tolerance=10000
\TITLEPAGE
\ITP{59}
\PUPT{1647}
\TITLE{\bf Non-Gaussian Isocurvature Perturbations}
\TITLE{\bf From Goldstone Modes Generated During Inflation}

\vskip 5pt 
{\baselineskip=14pt
\AUTHOR{Martin Bucher\footnote\dagger{Present Address: Institute 
for Theoretical
Physics, State University of New York, Stony Brook, NY 11794-3840. 
bucher@insti.physics.sunysb.edu}${}_{1,2}$ and Yong 
Zhu\footnote\ddagger{zhu@puhep1.princeton.edu}${}_{1}$}
\vskip 6pt
\PRINCETON
\vskip 6pt
\SB
}

\nobreak
\ABSTRACT{
\baselineskip 14pt 
We investigate non-Gaussian isocurvature perturbations
generated by the evolution of Goldstone modes during inflation. If a
global symmetry is broken {\it before} inflation, the resulting
Goldstone modes are disordered during inflation in a precise and 
predictable way.
After inflation these Goldstone modes order themselves in a self-similar
way, much as Goldstone modes in field ordering scenarios based on the
Kibble mechanism.  For $(H_{inf}^2/M_{pl}^2)\sim 10^{-6},$ through their 
gravitational interaction these Goldstone modes generate density
perturbations of approximately the right magnitude to explain the 
cosmic microwave background (CMB) anisotropy and seed the structure
seen in the universe today. We point out that for the pattern of 
symmetry breaking in which a global $U(1)$ is completely broken,
the inflationary evolution of the Goldstone field may be treated 
as that of a massless scalar field. Unlike the more commonly 
discussed case in which a global $U(1)$ 
is completely broken in a cosmological
phase transition,
in the inflationary case the production of defects can be made
exponentially small, so that Goldstone field evolution is 
completely linear. In such a model non-Gaussian perturbations
result because to lowest order density perturbations are 
sourced by products of Gaussian fields. Consequently, in this
non-Gaussian model N-point correlations may be calculated by
evaluating Feynman diagrams. We explore the issue of phase
dispersion and conclude that this non-Gaussian model 
predicts Doppler peaks in the CMB anisotropy. 
}

\rightline{[October 1996]}

\ENDTITLEPAGE

\REF\guth{A. Guth, ``Inflationary Universe: A Possible Solution to the Horizon 
and Flatness Problems," Phys. Rev. {\bf D23,} 347 (1981); A. Linde, ``A New 
Inflationary Universe Scenario: A Possible Solution of the
Horizon, Flatness, Homogeneity, Isotropy, and Primordial Monopole Problems,"
Phys. Lett. {\bf 108B,} 389 (1982); A. Albrecht and P. Steinhardt, ``Cosmology 
for Grand Unified Theories with Radiatively Induced Symmetry Breaking," 
Phys. Rev. Lett. {\bf 48,} 1220 (1982); A. Linde, ``Chaotic Inflation,''
Phys. Lett. {\bf 129B,} 177 (1983).}

\REF\iperth{S. Hawking, ``The Development of Irregularities in a Single
 Bubble Inflationary Universe,'' Phys. Lett. {\bf 115B}, 295 (1982);
A.A. Starobinsky, ``Dynamics of Phase Transition in the
New Inflationary Scenario and Generation of Perturbations,''
Phys. Lett. {\bf 117B}, 175 (1982);
A.H. Guth and S.-Y. Pi, ``Fluctuations in the New Inflationary
Universe,'' Phys. Rev. Lett. {\bf 49}, 1110 (1982);
J. Bardeen, P. Steinhardt and M. Turner, ``Spontaneous Creation of
Almost Scale-Free Density Perturbations in an Inflationary Universe,"
Phys. Rev. {\bf D28,} 679 (1983).}

\REF\wise{T.J. Allen, B. Grinstein and M.B. Wise, ``Non-Gaussian
Density Perturbations in Inflationary Cosmologies," Phys. Lett.
{\bf B197,} 66 (1987).}

\REF\axion{
M. Turner, F. Wilczek and A. Zee, ``Formation of Structure in an 
Axion Dominated Universe," Phys. Lett. {\bf 125B,} 35 (1983);
M. Axenides, R. Brandenberger and M. Turner, ``Development of Axion
Perturbations in an Axion Dominated Universe," Phys. Lett. {\bf 126B,}
178 (1983);
P. Steinhardt and M. Turner, ``Saving the Invisible Axion," Phys.
Lett. {\bf 129B,} 51 (1983);
A. Linde, ``Generation of Isothermal Density Perturbations in 
the Inflationary Universe," JETP Lett. {\bf 40,} 1333 (1984);
A. Linde, ``Generation of Isothermal Density Perturbations in 
the Inflationary Universe," Phys. Lett. {\bf 158B,} 375 (1985); 
D. Seckel and M. Turner, ``Isothermal Density Perturbations 
in an Axion Dominated Inflationary Universe," Phys. Rev. {\bf D32,} 
3178 (1985);
L. Kofman, ``What Initial Perturbations May Be Generated in Inflationary
Cosmological Models," Phys. Lett. {\bf 173B,} 400 (1986);
A. Linde and D. Lyth, ``Axionic Domain Wall Production During Inflation,"
Phys. Lett. {\bf 246B,} 353 (1990); 
M. Turner and F. Wilczek, ``Inflationary Axion Cosmology," Phys. Rev.
Lett. {\bf 66,} 5 (1991);
A. Linde, ``Axions in Inflationary
Cosmology," Phys. Lett. {\bf 259B,} 38 (1991);
D. Lyth, ``Axions and Inflation: Sitting in the Vacuum," Phys. 
Rev. {\bf D45,} 3394 (1992).}

\REF\salopek{ D.S. Salopek, J.R. Bond and J.M. Bardeen, ``Designing 
Density Fluctuation Spectra in Inflation," Phys. Rev. {\bf D40,}
1753 (1992); D.S. Salopek, ``Cold-Dark Matter Cosmology with 
Non-Gaussian Fluctuations from Inflation," Phys. Rev. {\bf D45,}
1139 (1992); D.S. Salopek, ``Can Non-Gaussian Fluctuations For 
Structure Formation Arise From Inflation," in T. Shanks et al., 
Eds., {\it Observational Tests of Cosmological Inflation,} (Boston: 
Kluwer Academic Publishers, 1991); 
D.S. Salopek, ``Non-Gaussian Microwave Background Fluctuations from 
Nonlinear Gravitational Effects,"  in G. Kunstatter et al., Eds.,
{\it Proceedings of the Fourth 
Canadian Conference on General Relativity and Relativistic 
Astrophysics,} (Singapore: World Scientific Publishing, 1992).}

\REF\fbardeen{Z.H. Fan and J.M. Bardeen, ``Predictions of a 
Non-Gaussian Model for Large Scale Structure," UW Preprint (1992).}

\REF\exinf{D. La and P.J. Steinhardt, ``Extended Inflationary
Cosmology," Phys. Rev. Lett. {\bf 62,}
376 (1989); Erratum: Ibid. {\bf 62,} 1066 (1989).}

\REF\top{
L. Kofman and A. Linde,  ``Generation of Density 
Perturbations in Inflationary Cosmology," Nucl. Phys. {\bf B282,}
555 (1987); 
E. Vishniac, K. Olive and D. Seckel, ``Cosmic Strings and
Inflation," Nucl. Phys. {\bf B289,} 717 (1987);
L. Kofman, A. Linde and J. Einasto, ``Cosmic Bubbles as Remnants 
from Inflation," Nature {\bf 326,} 48 (1987);
L. Kofman and D. Pogosyan, ``Nonflat Perturbations in Inflationary
Cosmology," Phys. Lett. {\bf 214B,} 508 (1988);
H. Hodges, G. Blumenthal, L. Kofman and J. Primack, ``Non-Standard
Primordial Perturbations from a Ploynomial Inflaton Potential,"
Nucl. Phys. {\bf B335,} 197 (1990);
D. Lyth, ``Inflationary Energy Scales, Cosmic Strings and Textures,"
Phys. Lett. {246B,} 359 (1990);
L. Kofman, G. Blumenthal, H. Hodges and J. Primack, ``Generation 
of Nonflat and Non-Gaussian Perturbations from Inflation," in 
D. Latham and L. Da Costa, Eds., {\it Large Scale Structures and Peculiar
Motion in the Universe,} (San Francisco: APS Conference Series)
(1990);
H. Hodges and J. Primack, ``Strings, Textures and Inflation,"
Phys. Rev. {\bf D43,} 3155 (1991);
L. Kofman, ``Primordial Perturbations From Inflation," Physica Scripta
{\bf T36,} 108 (1991);
A. Linde, ``Strings, Textures, Inflation, and Spectrum Bending,"
Phys. Lett {\bf 284B,} 215 (1992);
N. Turok and Y. Zhu, ``Inflationary Textures," in preparation.
}

\REF\topb{R. Basu, A. Guth and A. Vilenkin, ``Quantum
Creation of Topological Defects During Inflation," Phys. Rev.
{\bf D44,} 340 (1991).}

\REF\avone{A. Vilenkin, ``Cosmological Density Perturbations Produced by
a Goldstone Field," Phys. Rev. Lett. {\bf 48,} 59 (1982).}

\REF\turok{N. Turok and D. Spergel, ``Scaling Solution for Cosmological
$\sigma $ Models at Large N," Phys. Rev. Lett. {\bf 66,} 3093 (1991).}

\REF\schmid{H. M\"uller and C. Schmid, ``Energy Fluctuations
Generated by Inflation," (Jan. 1994) (gr-qc 9401020);
H. M\"uller and C. Schmid, ``Non-Gaussian Primordial Fluctuations
in Inflationary Cosmology," (Oct. 1994) (gr-qc 9412022);
H. M\"uller and C. Schmid, ``Energy Density Fluctuations in 
Inflationary Cosmology," (Dec. 1994) (gr-qc 9412021).}

\REF\trk{N. Turok,
``A Causal Source Which Mimics Inflation," (astro-ph 9607109);
N. Turok, ``Subdegree Scale Microwave Anisotropies From Cosmic Defects,"
(astro-ph 9606087).}
\REF\ct{R. Crittenden and N. Turok, ``The Doppler Peaks from Cosmic Texture,"
Phys. Rev. Lett. {\bf 75,} 2642 (1995) (astro-ph 9505120).}
\REF\durr{R. Durrer, A. Gangui and M. Sakellariadou, ``Doppler 
Peaks: A Fingerprint
of Topological Defects," Phys. Rev. Lett. {\bf 76,} 579 (1996).}
\REF\mafc{J. Magueijo, A. Albrecht, P. Ferreira 
and D. Coulson, ``The Structure
of Doppler Peaks Induced By Active Perturbations," Phys. Rev. {\bf D54,}
3727, (1996) (astro-ph 9605047);
J. Magueijo, A. Albrecht, D. Coulson and  P. Ferreira,
``Doppler Peaks from Active Perturbations," Phys. Rev. Lett. {\bf 76,}
2617 (1996).}
\REF\hsw{
W. Hu, D. Spergel and M. White, ``Distinguishing Causal Seeds from Inflation,"
(astro-ph 9605193).}

\REF\stoc{See, for example, A. Linde,  D. Linde and
A. Mezhlumian, ``From the Big Bang Theory to the Theory
of a Stationary Universe," Phys. Rev. {\bf D49,} 1783 (1993);
A. Linde, {\it Particle Physics and Cosmology,}   
(Switzerland: Harwood, 1992), and references therein.}

\REF\cpma{See, for example, 
C.P. Ma and E. Bertschinger, ``Cosmological Perturbation Theory
in Synchronous and Conformal Newtonian Gauges," Ap. J. {\bf 455,} 7 (1995).
(astro-ph 9506072).}

\REF\mukhanov{A. Linde and V. Mukhanov, ``Non-Gaussian Isocuravture
Perturbations from Inflation," (astro-ph 9610219) (1996).}

\baselineskip=18pt
\def\H{{\cal H}}
\def\k{{\bf k}}
\def\x{{\bf x}}

\def\gtorder{\mathrel{\raise.3ex\hbox{$>$}\mkern-14mu
             \lower0.6ex\hbox{$\sim$}}}
\def\ltorder{\mathrel{\raise.3ex\hbox{$<$}\mkern-14mu
             \lower0.6ex\hbox{$\sim$}}}

\chapter{Introduction}

A key problem of modern cosmology is the origin
of the primordial fluctuations that later led to 
structure in the universe. From a theoretical perspective,
at this point the most satisfying theory for the very early universe is 
inflationary cosmology because of its nice resolution of the horizon, 
smoothness, flatness, and monopole problems.\refmark{\guth } In its 
simplest form, 
inflation with a single real scalar field predicts an approximately 
scale-invariant spectrum of Gaussian adiabatic primordial density 
perturbations.\refmark{\iperth } But if other light scalar fields 
are present during 
inflation (light relative to the expansion rate $H_{inf}$ during inflation), 
other types of density perturbations are generated as well, and these other 
types of perturbations leave a different and distinguishable imprint 
on the universe today. 

The observation that inflation can produce perturbations other
than the commonly discussed almost scale-free Gaussian 
adiabatic density perturbations is not new. Possibilities have
been pointed out by various authors. In inflationary models 
with a Peccei-Quinn symmetry broken before inflation,
the inflationary expansion disorders the axion field, which
during inflation may be regarded as a massless scalar field.
After inflation, when the axion mass becomes relevant, 
fluctuations in the axion field translate into perturbations 
in the axion mass density. Allen, Grinstein, and 
Wise \refmark{\wise } observed that a fourth-order derivative 
term in the effective theory for the Peccei-Quinn symmetry breaking
gives rise to non-Gaussian fluctuations in the axion density.
There are other possibilities for non-Gaussianity in axion density
perturbations from inflation;\refmark{\axion } however, in 
many cases inflationary axion perturbations are very nearly
Gaussian. It has also been pointed out that for inflation with
multiple scalar fields possibilities exist for non-Gaussian
fluctuations.\refmark{\salopek, \fbardeen } For example, 
when the classical
slow-roll path bifurcates into two paths toward the minimum, 
in what may be described as hitting a mogul, quantum 
fluctuations about the slow-roll path determine which
side of the mogul the inflaton field chooses and non-Gaussian
fluctuations result. Fluctuations from finite size bubbles
in extended inflation\refmark{\exinf }
give non-Gaussian density perturbations. 
Also by combining inflation with various sorts of topological
defect models non-Gaussian perturbations may 
be produced.\refmark{\top ,\topb }

In this paper we investigate the evolution of Goldstone modes during
inflation and the density perturbations subsequently generated.
We assume an exactly massless Goldstone mode 
and also that the potential for the inflaton field is such 
that the usual adiabatic Gaussian perturbations from 
quantum fluctuations of the inflaton field are so small as
to be observationally irrelevant. We assume that unlike 
in axion models the global symmetry giving rise to the Goldstone 
mode remains exact today. The effect of even a small Goldstone mass 
drastically changes the analysis of the perturbations. Vilenkin 
in 1982 pointed out how the ordering of an initially disordered 
Goldstone field can generate density perturbations of the right 
order to explain the origin of structure in the 
universe.\refmark{\avone } Vilenkin considered a Goldstone field
initially disordered through a symmetry-breaking phase transition,
so that on large scale there are no correlations. The evolution 
of Goldstone modes for large N have been studied by Turok and
Spergel.\refmark{\turok } Here we 
assume that the global symmetry is broken {\it before} inflation
so that the inflationary dynamics of the Goldstone fields
determine the state of the Goldstone field on superhorizon 
scales at the end of inflation. 

M\"uller and Schmid have also considered density fluctuations
arising from an inflationary Goldstone mode.\refmark{\schmid }
Their analysis differs from ours in that they compute 
only correlations of the Goldstone field contribution to the
stress-energy. The cosmological perturbations that we observe
today are perturbations in the gravitational potential and in 
the total stress-energy. Since the contribution of the Goldstone 
modes to the total stress-energy is small, it is justified to
ignore the effect of gravity on the evolution of the Goldstone 
modes, treating them as a stiff source. The rest of the matter 
in the universe, which gives the overwhelming contribution to
the total stress-energy, may be treated as a single-component perfect
fluid, first with the equation of state of radiation and later
with a pressureless equation of state. The Goldstone stiff source 
excites perturbations of this single-component fluid coupled
to gravity. Cosmological perturbations observed today are 
computed using Green functions---their correlations 
are related by an integral transform to correlations at 
earlier times of the sort considered by M\"uller and Schmid.

Inflationary Goldstone modes are interesting for two reasons. First, 
if we are to confront inflation with observation, we must fully explore 
the possibilities offered by inflationary cosmology rather than 
exclusively focusing on the predictions of the simplest models. The second 
motivation has to do with the simplicity of this model and its qualitative 
similarity to field ordering models based on the Kibble mechanism (i.e., 
cosmic strings, global monopoles, textures, nonlinear Goldstone modes, etc.). 
For inflationary Goldstone modes, as we shall show, in many situations 
their evolution is linear, or almost linear. 
The resulting density fluctuations, however, are non-Gaussian
because they are sourced by the stress-energy
from the Goldstone modes, which is the sum of products of two Gaussian 
fields rather than a Gaussian field itself. Because of this, the 
expectation values of the
CMB multipole moments are four-point functions in 
a Gaussian theory, and thus comparatively simple to evaluate. Despite 
the underlying Gaussian character of these linear inflationary modes, this 
model is remarkably similar to field ordering models based on the Kibble
mechanism. In both cases one starts (either after inflation or after
the symmetry-breaking phase transition) with a disordered order parameter
field that orders itself in a self-similar way, with the characteristic
length scale at any time equal to the Hubble length at that time. In both
cases the matter density perturbations are sourced by a Goldstone mode 
stress-energy $\Theta _{\mu \nu }({\bf x},t)$ that is non-Gaussian, so that a
simple decomposition into momentum modes is not possible. The
principal difference between inflationary Goldstone modes and 
Goldstone modes from the Kibble mechanism lies in the correlations  
of the initial field configurations on superhorizon scales. 
With the Kibble mechanism there are initially
no correlations on large scales because of considerations of 
causality, but in the inflationary case at reheating there exist correlations 
on arbitrarily large scales. However, this difference is expected to have 
a small effect, probably only changing the parameters of a scaling 
solution but not its qualitative behavior. 
 
Recently there has been much discussion in the literature about the 
structure of the small-angle CMB multipole moments in field ordering 
theories.\refmark{\trk -\hsw } 
An important qualitative issue is whether field ordering
theories predict oscillations, or {\it Doppler} peaks, 
in the large-$\ell $ CMB multipole moments, similar to those predicted for 
adiabatic Gaussian fluctuations 
from inflation. For adiabatic Gaussian fluctuations each mode in momentum 
space has a definite fixed {\it phase}, although its real amplitude is a 
Gaussian random variable. Therefore 
modes of wavelength comparable to or smaller than the Hubble length at
last scattering undergo acoustic oscillations with a definite, fixed
phase. By contrast, for non-Gaussian isocurvature models these acoustic
modes have at least some dispersion in their phase. Since these
modes are sourced primarily during horizon crossing over a time of order
the Hubble time, there is no reason to expect what would be the decaying 
mode at earlier times to be absent, and since the source is nonlinear, 
there is no reason to expect that the time dependence of the source for a 
given mode,
as opposed to its overall amplitude, not to fluctuate. A fixed time dependence 
would eliminate any phase dispersion. Therefore, it is clear that oscillations
of the CMB moments in field ordering theories are at least
to some extent suppressed, but how much is a detailed quantitative question. 
While there definitely exist effects that 
tend to smear the phase, it is not {\it a priori} clear 
whether these effects are
strong enough to suppress the oscillatory character of the small-angle 
CMB multipole moments. Crittenden and Turok\refmark{\ct }
have found oscillatory Doppler 
peaks for global textures (where the order parameter space is 
$SU(2)\cong S^3).$ Their model for computing the CMB moments 
assumes a fixed time dependence for the source for each mode, thus not 
allowing for any phase dispersion. Durrer et al. \refmark{\durr }
have also studied the CMB moments for textures, obtaining results 
essentially in agreement with 
Crittenden and Turok. Albrecht et al.,\refmark{\mafc }
on the other hand, have suggested
that phase dispersion may be very significant, especially for cosmic 
string models, which they claim exhibit no secondary Doppler peaks. 
It is hoped that the CMB anisotropy moments
from linearized inflationary Goldstone modes, which 
because of its simplicity can be solved exactly in terms of four-point 
functions in a Gaussian theory, may shed some light on these questions.  

Goldstone modes arise through the breaking of a global symmetry $G$ to a 
smaller group $H,$ giving rise to massless modes equal in number to the 
difference in dimensionality of the two groups. Most studied has been the 
evolution of such modes after a cosmological phase transition. In these field
ordering or topological defect models, it is typically assumed that initially
(either because of preferred initial conditions or because of a prior
epoch of inflation) one starts with an exactly homogeneous and isotropic
universe at a high temperature in which the symmetry $G$ is unbroken. Later, as
the universe cools, a phase transition takes place in which the symmetry $G$ 
breaks to $H.$ Immediately after the phase transition at each point an 
orientation of $G/H$ is chosen at random, but beyond some correlation length,
which by considerations of causality cannot exceed the horizon size, the 
orientations are uncorrelated.    
Subsequently the order parameter field orders itself in a self-similar way,
described by a scaling solution, which
in general must be determined numerically. 

In this paper
we consider a situation in which the symmetry $G$ is broken to $H$
{\it before} inflation. In this case the state of the field $G/H$ at 
the end of inflation is completely determined by the inflationary dynamics
of the field $G/H$ during inflation. All vestiges of initial conditions
before inflation are erased. After inflation the coset field evolves 
classically, becoming aligned on increasingly larger scales, just as in the 
scenarios based on the Kibble mechanism. Although the field ordering dynamics 
in the two cases (after the end of inflation or after the phase transition)
are the same, the initially conditions at the respective times are quite
different. After the phase transition the coset field is uncorrelated
over large distances---one starts with a white noise spectrum on large scales. 
By contrast, at the end of inflation there exist correlations on 
arbitrarily large scales.

In the case where $H_{inf}\ll f_g$ (where $H_{inf}$ is the Hubble 
constant during
inflation and $f_g$ is the symmetry breaking scale, roughly the radius
of the coset space), one may adopt the following stochastic picture of
the inflationary dynamics of the coset field, taken from the stochastic
approach to chaotic inflation.\refmark{\stoc }
We shall ignore fluctuations of the coset
field on scales less than the Hubble length $H_{inf}^{-1},$ treating the 
coset field as classical and constant over a Hubble volume, a cube 
of volume $H_{inf}^{-3}.$ As the universe expands by a factor of two, each
such cube subdivides into eight subcubes, each of the original size, and in 
each 
of these subcubes the coset field takes random step of order $\Delta \phi
\approx H_{inf}.$ 
The random steps in different subcubes are uncorrelated. This process
repeats until the end of inflation. Physically, the random steps represent the 
freezing in of quantum fluctuations on subhorizon scales by the 
inflationary expansion. 

In the most general case, owing to nonlinearity, the evolution 
of the nonlinear sigma 
model $G/H$ during and after inflation is quite complicated and thus not
amenable to an analytic treatment. However in special cases nonlinearity 
plays a negligible role. When $G/H\cong S^1,$ the nontrivial topology of 
$S^1$ is irrelevant, and we may replace $S^1$ by the real line, treating 
the Goldstone mode as a free massless real scalar field. If the potential
in the radial direction is sufficiently stiff, the process by which the
field jumps over the origin creating a configuration of nonzero winding
number is exponentially suppressed.\refmark{\topb }
This process may be described as the 
nucleation of a cosmic string loop. For the loop not to recollapse,
it must have a radius larger than a Hubble length. Note that for this
pattern of symmetry breaking, in scenarios based on the Kibble mechanism
nonlinearity is always important, because there is no way to suppress
the formation of cosmic strings. The other case in which inflationary
Goldstone modes may be linearized arises when $H_{inf}\ll f_g.$ In this case,
provided one does not consider too large a volume, the Goldstone
field never wanders far enough from its average value for nonlinear 
effects to become important. 

The organization of this paper is as follows. In section II we 
present the initial condition for a linearized Goldstone mode at
the end of inflation and give its evolution into 
the radiation-dominated era as well as the resulting 
contribution of the Goldstone mode to the stress-energy.  
In section III the resulting stress-energy is renormalized
and the effect of this source through gravity on the radiation 
fluid is computed. A derivation of the Green's functions
for computing the perturbations of the radiation fluid as
an integral transform over the stiff source stress-energy
is relegated to an appendix.
Because our computation does not include
the decaying modes of the Goldstone field, 
our calculations become unreliable on
subhorizon scales. We follow the evolution of the radiation
fluid modes until horizon crossing and compute the phase
dispersion in terms of a covariance matrix of two point functions 
of the two radiation fluid modes. We find the amount of phase 
dispersion to be small. In Section IV we present some
concluding remarks indicating future directions.

\chapter{Goldstone Mode Evolution and Stress-Energy}

For simplicity we consider a single Goldstone mode represented
by a real massless scalar field, as in the $G/H\cong S^1$ model.
The generalization to several Goldstone modes is straightforward.
After the end of inflation we may treat this field as a Gaussian
ensemble of classical fields that evolve classically, just as 
one typically assumes in the discussion of the usual adiabatic 
perturbations from inflation. Initially, on superhorizon scales 
one has
$$\phi ({\bf x},\eta =0)=
{H_{inf}\over \sqrt{2}}\int {d^3k\over (2\pi )^3}~
k^{-3/2}a({\bf k})~e^{i{\bf k}\cdot {\bf x}}
\eqn\bba$$
where $H_{inf}$ is the Hubble constant during inflation and
the $a({\bf k})$'s are complex Gaussian random variables subject to the 
constraint $a({\bf k})=a(-{\bf k})^*$ with the two-point function
$$\langle a({\bf k})a({\bf k}')\rangle =(2\pi )^3\cdot 
\delta ^3({\bf k}+{\bf k}'). 
\eqn\bbb$$ 
The subsequent evolution of the Goldstone modes is given by 
$$\phi ({\bf x},\eta )=
{H_{inf}\over \sqrt{2}}\int {d^3k\over (2\pi )^3}~
k^{-3/2}a({\bf k})~e^{i{\bf k}\cdot {\bf x}}~
T(\eta ,k)
\eqn\bbc$$
where $T(\eta ,k)$ satisfies 
$$\ddot T+{2\dot a(\eta )\over a(\eta )}\dot T+k^2T=0
\eqn\bbd$$
and the boundary conditions $T(\eta =0,k)=1,$ $\dot T(\eta =0,k)=0.$
Here the spacetime metric is $ds^2=a^2(\eta )\cdot [-d\eta ^2
+d{\bf x}^2 ]$ and the dots denote derivatives with respect to
conformal time $\eta .$

The stress-energy from the massless scalar field is given by
$$\Theta _{\mu \nu }=(\partial _\mu \phi )(\partial _\nu \phi )
-{1\over 2}g_{\mu \nu }[g^{\alpha \beta }
(\partial _\alpha \phi )(\partial _\beta \phi )].
\eqn\bbe$$

Assuming a radiation-dominated universe so that $a(\eta )=\eta ,$
we may solve eqn. \bbd ~ explicity, obtaining
$$T(\eta ;k)=j_0(k\eta )={\sin [k\eta ]\over (k\eta )}.
\eqn\bbf$$

Here we have given an explicit analytical solution for the evolution
of the stiff source. In most field ordering theories (such as
textures, global monopoles, and cosmic strings) the evolution of the
stiff source must be computed numerically on a huge lattice.

In computing matrix elements involving $\Theta _{\mu \nu }({\bf k})$ it is
necessary to renormalize. In this paper we are concerned with 
computing two-point functions of the form 
$$\langle \Theta _{\mu \nu }({\bf k})\Theta _{\mu '\nu '}(-{\bf k})\rangle .
\eqn\boo$$
These can be renormalized \'a la Zimmerman by 
replacing the unrenormalized integrand $I_{unrenorm}({\bf k})$
with $I_{renorm}({\bf k})=I_{unrenorm}({\bf k})-I_{unrenorm}({\bf k}=0).$
Given the approximations used here, in which the decaying mode is ignored 
resulting in a description for the behavior of the scalar field 
accurate only on superhorizon scales, the need to renormalize is not 
completely manifest, because
integrating over the unrenormalized integrands gives convergent integrals,
with dominant contribution from momenta of order $k\approx H_{inf}.$ 
But $k\approx H_{inf}$ is precisely where the approximations used here 
break down. The ultraviolet divergence has been avoided because we 
have extrapolated the infrared enhanced correlations on superhorizon 
scales to subhorizon scales. In a sense through our approximations we 
have introduced a cutoff 
at $k_{internal}\approx H_{inf}.$ 
The renormalized matrix elements can be computed straightforwardly by
subtracting the integrand at $k=0.$ After this subtraction 
at any given time $\eta $ the dominant
modes contributing to $\Theta _{\mu \nu }$ are peaked around 
$k\approx \eta ^{-1}$ (i.e., of wavelength of order the horizon
size, or Hubble length), as one might expect.

For future reference we give the expansion of the Fourier
components 
$$ \Theta _{\mu \nu }(\k )=\int d^3x~e^{-i\k \cdot \x }~ 
\Theta _{\mu \nu }(\x ) \eqn\bbg$$
in terms of the Gaussian random variables $a(\k )$ and decompose these 
Fourier components into their {\it scalar,} {\it vector,} 
and {\it tensor} parts.

Substituting eqn. \bbc ~ into eqn. \bbe , one obtains
$$\eqalign{
&\Theta _{00}(\k )={H_{inf}^2\over 2}\int {d^3k'\over (2\pi )^3}~
{a(\k -\k ')a(\k ')\over 
\vert \k -\k '\vert ^{3\over 2}~\vert \k '\vert ^{3\over 2}}\cr
&~~~~\times \left[ {1\over 2}
{\partial T(\eta ,\vert \k -\k '\vert )\over \partial \eta ~~~}
{\partial T(\eta ,\vert \k '\vert )\over \partial \eta ~~~}
-{1\over 2}(\k -\k ')\cdot \k ' ~
T(\eta ,\vert \k -\k '\vert )~T(\eta ,\vert \k '\vert )
\right] ,\cr }
\eqn\bbh$$
$$\eqalign{
&\Theta _{0i}(\k )={1\over 4}H_{inf}^2\int {d^3k'\over (2\pi )^3}~
{a(\k -\k ')a(\k ')\over 
\vert \k -\k '\vert ^{3\over 2}~\vert \k '\vert ^{3\over 2}}\cr 
&~~~~\times \left[  
+i~(\k ')_i
{\partial T(\eta ,\vert \k - \k '\vert )\over \partial \eta ~~~}~
T(\eta ,\vert \k '\vert )
+i~(\k-\k ')_i
{\partial T(\eta ,\vert \k '\vert )\over \partial \eta ~~~}~
T(\eta ,\vert \k-\k '\vert )
\right] ,\cr }
\eqn\bbi$$
$$\eqalign{
&\Theta _{ij}(\k )={H_{inf}^2\over 2}\int {d^3k'\over (2\pi )^3}~
{a(\k -\k ')a(\k ')\over
\vert \k -\k '\vert ^{3\over 2}~\vert \k '\vert ^{3\over 2}}\cr
&~~\times \biggl[ {1\over 2}
{\partial T(\eta ,\vert \k -\k '\vert )\over \partial \eta ~~~}
{\partial T(\eta ,\vert \k '\vert )\over \partial \eta ~~~} ~\delta _{ij}
+{1\over 2}(\k -\k ')\cdot \k ' ~
T(\eta ,\vert \k -\k '\vert )~T(\eta ,\vert \k '\vert )~\delta _{ij}\cr 
&~~~~~~~~~~~~~~
-(\k -\k ')_i~(\k ')_j~T(\eta ,\vert \k -\k '\vert )~T(\eta ,\vert \k '\vert )
\biggr] .\cr }
\eqn\bbh$$

We decompose the spatial-spatial part of $\Theta _{ij}(\k )$ as follows:
$$
\Theta _{ij}^S(\k )=\delta _{ij}~\Theta _L(\k )+
\left( {k_ik_j-{1\over 3}k^2\delta _{ij}\over k^2}\right) 
\Theta _T(\k ),
\eqn\bbi$$
so that
$$\eqalign{
&\Theta _L(k)
={H_{inf}^2\over 2}\int {d^3k'\over (2\pi )^3}~{a(\k -\k ')a(\k ')\over
\vert \k -\k '\vert ^{3\over 2}~\vert \k '\vert ^{3\over 2}}\cr
&~~\times \biggl[ {1\over 2}
{\partial T(\eta ,\vert \k -\k '\vert )\over \partial \eta ~~~}
{\partial T(\eta ,\vert \k '\vert )\over \partial \eta ~~~} 
+{1\over 6}(\k -\k ')\cdot \k ' ~
T(\eta ,\vert \k -\k '\vert )~T(\eta ,\vert \k '\vert )~\biggr] \cr }
\eqn\bbj$$
and 
$$\eqalign{
&\Theta _T(k)
={H_{inf}^2\over 2}\int {d^3k'\over (2\pi )^3}~{a(\k -\k ')a(\k ')\over
\vert \k -\k '\vert ^{3\over 2}~\vert \k '\vert ^{3\over 2}}\cr
&~~\times \biggl[ 
\left( -(\k \cdot \k ')+{3\over 2}{(\k \cdot \k ')^2\over k^2}
-{1\over 2}~\k ^{\prime ~2} \right) 
T(\eta ,\vert \k -\k '\vert )~T(\eta ,\vert \k '\vert )~\biggr] .\cr }
\eqn\bbj$$

\chapter{Scalar Density Perturbations in a Radiation-Dominated Universe}

In the previous section we computed the evolution of the stiff
source (i.e., the linearized Goldstone modes) and the stress-energy
source $\Theta _{\mu \nu }$ resulting from these linearized Golstone
modes. In this section we compute the {\it scalar} density perturbations
generated by this source, relegating many of the computational
details and a precise description of the notational conventions
to Appendix A. 

For computational simplicity, to make the problem as scale free as 
possible, we assume a radiation-dominated universe, with $a(\eta )
=\eta $ and a single-component fluid with $c_s^2={1\over 3}$ and 
work in momentum space. It follows using the Green functions derived
in Appendix A that for the Newtonian potential
$$  
\psi (\k ,\eta )= \eta ^{-1}\cdot \left[ ~
\psi (\k ,j,\eta )~j_1(k\eta /\sqrt{3})+
\psi (\k ,y,\eta )~y_1(k\eta /\sqrt{3})~\right] ,\eqn\xxb$$
where 
$$\eqalign{
\psi (\k ,\eta ,j)&=-(4\pi G)~\left({k\over \sqrt{3}}\right) 
\int _0^\eta d\bar \eta ~\bar \eta ^{3}~
y_1(k\bar \eta /\sqrt{3})\cr &~
\times \left[
\Theta _L(\k ,\bar \eta )-{1\over 3}~\Theta _{00}(\k ,\bar \eta )+
{2\over 3}
\Theta _T(\k ,\bar \eta )-
{2\dot \Theta _T(\k ,\bar \eta )\over k^2\bar \eta }
\right] \cr }\eqn\xxb$$
and similarly
$$\eqalign{
\psi (\k ,\eta ,y)&=+(4\pi G)~\left({k\over \sqrt{3}}\right) 
\int _0^\eta d\bar \eta ~\bar \eta ^{3}~
j_1(k\bar \eta /\sqrt{3})\cr &~
\times \left[
\Theta _L(\k ,\bar \eta )-{1\over 3}\Theta _{00}(\k ,\bar \eta )+
{2\over 3}
\Theta _T(\k ,\bar \eta )-
{2\dot \Theta _T(\k ,\bar \eta )\over k^2\bar \eta }
\right] .\cr }\eqn\xxc$$

We now formulate in more quantitative terms the issue of phase 
dispersion discussed in the introduction. We compute the 
covariance matrix
$$ {\rm Cov}(\k ,\eta )= \pmatrix{ 
\langle \psi (\k ,\eta ,j) ~\psi (-\k ,\eta ,j)\rangle &
\langle \psi (\k ,\eta ,j) ~\psi (-\k ,\eta ,y)\rangle \cr 
\langle \psi (\k ,\eta ,y) ~\psi (-\k ,\eta ,j)\rangle &
\langle \psi (\k ,\eta ,y) ~\psi (-\k ,\eta ,y)\rangle \cr 
} .\eqn\xxd$$
Here $j$ and $y$ refer
to the two types of Bessel functions---or equivalently, the growing
and decaying modes, respectively. 

The eigenvalues of the covariance matrix provide a measure of the phase
dispersion. For the usual adiabatic perturbations from inflation (for
which only a growing mode is present), the covariance matrix is 
proportional to $\pmatrix{ 1&0\cr 0&0\cr }.$ In the most general case
with no phase dispersion, this matrix has one vanishing eigenvalue,
indicating a definite fixed linear relationship between $\psi (\k ,j)$
and $\psi (\k ,y).$ By contrast, a completely random phase gives
a covariance matrix proportional to $\pmatrix{ 1&0\cr 0&1\cr }.$

We now compute the matrix element
$$\eqalign{
&\langle \psi (\k ,\eta ,y)~\psi (-\k ',\eta ',y)\rangle  \cr 
&=H_{inf}^4(4\pi G)^2 {kk'\over 12}\cdot \int {d^3k_1\over (2\pi )^3}~
\int {d^3k_2\over (2\pi )^3}~
\int _0^\eta  d\eta _1~\eta _1^3 ~
\int _0^{\eta '}d\eta _2~\eta _2^3 ~
j_1(k\eta _1/\sqrt{3})~j_1(k\eta _2/\sqrt{3}) \cr 
&\times {
\langle \hat a(\k -\k _1)~\hat a(\k _1)~\hat a(-\k '+\k _2)~\hat a(-\k _2)
\rangle \over 
\vert \k -\k _1\vert ^{3/2}\vert \k _1\vert ^{3/2}
\vert \k '-\k _2\vert ^{3/2}\vert \k _2\vert ^{3/2} }\cr 
&\times \Biggl[
{1\over 3}{\partial j_0(\vert \k -\k _1\vert \eta _1)\over \partial \eta _1~~~}
{\partial j_0(k_1\eta _1)\over \partial \eta _1~~~}\cr
&~~~
+\left\{ {(\k \cdot \k _1)^2\over k^2}-{(\k \cdot \k _1)\over 3}-{2\over 3}k_1^2\right\} 
j_0(\vert \k -\k _1\vert \eta _1)~j_0(k_1\eta _1)\cr 
&~~~~~~ +{2\over k^2}\left\{ {1\over 2}k_1^2+(\k \cdot \k _1)-
{3\over 2} {(\k \cdot \k _1)^2\over k^2}\right\} {1\over \eta _1}
{~\partial \over \partial \eta _1}\biggl\{ 
j_0(\vert \k -\k _1\vert \eta _1)~j_0(k_1\eta _1)
\biggr\} \Biggr] \cr 
&\times \Biggl[
{1\over 3}{\partial j_0(\vert \k '-\k _2\vert \eta _2)\over \partial \eta _2~~~}
{\partial j_0(k_2\eta _2)\over \partial \eta _2~~~}\cr 
&~~~
+\left\{ {(\k '\cdot \k _2)^2\over k^{\prime 2}}-{(\k '\cdot \k _2)\over 3}
-{2\over 3}k_2^2\right\} 
j_0(\vert \k '-\k _2\vert \eta _2)~j_0(k_2\eta _2)\cr
&~~~~~~ +{2\over k^{\prime 2}}\left\{ {1\over 2}k_2^2+(\k '\cdot \k _2)-
{3\over 2} {(\k '\cdot \k _2)^2\over k^{\prime 2}}\right\} {1\over \eta _2}
{~\partial \over \partial \eta _2}\biggl\{ 
j_0(\vert \k '-\k _2\vert \eta _2)~j_0(k_2\eta _2)
\biggr\} \Biggr] \cr
&~~~-\Biggl( {\rm integrand~with~} \Theta _{\mu \nu }(\k )
\to \Theta _{\mu \nu }(\k =0),~\Theta _{\mu \nu }(\k ')
\to \Theta _{\mu \nu }(\k '=0)\Biggr) .\cr }\eqn\xaa$$
The matrix element
$$\langle a(\k -\k _1)~a(\k _1)~a(-\k '+\k _2)~a(-\k _2) \rangle
\eqn\bbb$$
as a consequence of Wick's theorem may be decomposed into a sum of products
of two-point functions with three terms, one of which vanishes, specifically
the one that would result from computing
$\langle \psi (\k ,\eta ,j)\rangle ~~\langle \psi (-\k ',\eta ,j) \rangle $.
Therefore eqn. \bbb ~ equals
$$(2\pi )^6\cdot \delta ^3(\k -\k ')\cdot \biggl[ \delta ^3(\k _1-\k _2)+
\delta ^3(\k-\k _1-\k _2)\biggr] ,\eqn\bbc$$
allowing us to set $\k =\k '$ and to consider two cases: (1) $\k _2=\k _1,$
and (2) $\k _2=\k -\k _1.$ By a simple symmetry the two cases give the same
contribution, so that eqn. \xaa ~ becomes
$$\eqalign{
&M_{yy}(k;\eta _1,\eta _2)\cdot \delta ^3(\k -\k ')\cr 
&=H_{inf}^4(4\pi G)^2 {k^2\over 12}\cdot \delta ^3(\k -\k ')~\int d^3k_1~
\int _0^\eta d\eta _1~\eta _1^3 ~
\int _0^{\eta '}d\eta _2~\eta _2^3 ~
{j_1(k\eta _1/\sqrt{3})~j_1(k\eta _2/\sqrt{3})\over 
\vert \k -\k _1\vert ^3\vert \k _1\vert ^3} \cr
&\times \Biggl[
{1\over 3}{\partial j_0(\vert \k -\k _1\vert \eta _1)\over \partial \eta _1~~~}
{\partial j_0(k_1\eta _1)\over \partial \eta _1~~~}\cr 
&~~~
+\left\{ {(\k \cdot \k _1)^2\over k^2}-{\k \cdot \k _1\over 3}-
{2\over 3}k_1^2\right\}
j_0(\vert \k -\k _1\vert \eta _1)~j_0(k_1\eta _1)\cr
&~~~~~~ +{2\over k^2}\left\{ {1\over 2}k_1^2+(\k \cdot \k _1)-
{3\over 2} {(\k \cdot \k _1)^2\over k^2}\right\} {1\over \eta _1}
{~\partial \over \partial \eta _1}\biggl\{
j_0(\vert \k -\k _1\vert \eta _1)~j_0(k_1\eta _1)
\biggr\} \Biggr] \cr
&\times \Biggl[
{1\over 3}{\partial j_0(\vert \k -\k _1\vert \eta _2)\over \partial \eta _2~~~}
{\partial j_0(k_1\eta _2)\over \partial \eta _2~~~}\cr 
&~~~
+\left\{ {(\k \cdot \k _1)^2\over k^2}-{\k \cdot \k _1\over 3}-
{2\over 3}k_1^2\right\}
j_0(\vert \k'-\k _1\vert \eta _2)~j_0(k_1\eta _2)\cr
&~~~~~~ +{2\over k^2}\left\{ {1\over 2}k_1^2+(\k \cdot \k _1)-
{3\over 2} {(\k \cdot \k _1)^2\over k^{2}}\right\} {1\over \eta _2}
{~\partial \over \partial \eta _2}\biggl\{ j_0(\vert \k -
\k _1\vert \eta _2)~j_0(k_1\eta _2)
\biggr\} \Biggr] \cr
&~~~-\Biggl( {\rm integrand~with~} \Theta _{\mu \nu }(\k )
\to \Theta _{\mu \nu }(\k =0) \Biggr) .\cr }
\eqn\xab$$

In order to compare the matrix elements in the most meaningful way,
we consider the quantities
$$\eqalign{
F_{jj}(k,\eta )&=k^3\cdot \left[ {j_1(k\eta /\sqrt{3})\over \eta }\cdot 
{j_1(k\eta /\sqrt{3})\over \eta }\right] \cdot M_{jj}(k;\eta ,\eta ),\cr
F_{jy}(k,\eta )&=k^3\cdot \left[ {j_1(k\eta /\sqrt{3})\over \eta }\cdot 
{y_1(k\eta /\sqrt{3})\over \eta }\right] \cdot M_{jy}(k;\eta ,\eta ),\cr 
F_{yy}(k,\eta )&=k^3\cdot \left[ {y_1(k\eta /\sqrt{3})\over \eta }\cdot 
{y_1(k\eta /\sqrt{3})\over \eta }\right] \cdot M_{yy}(k;\eta ,\eta ).\cr }
\eqn\gbb$$
In Figure 1 the $F$ matrix elements are plotted for various $k$ at $\eta =1.$
For $k\eta \gtorder 1$ the decaying modes of the Goldstone field become
important, and our calculation here becomes unreliable because of the 
neglect of these modes. 
[Because of the scaling properties of a radiation-dominated universe, 
any equal time matrix elements may be
obtained by a trivial rescaling of equal time matrix elements at $\eta =1.$]
The $k^3$ factor has been introduced to make $F$ scale free---for a
Harrison-Zeldovich spectrum $F$ would be independent of $k.$ Instead
we oberve a linear dependence with $k$ in the $F$ matrix elements, 
reflecting the fact that the radiation fluid/scalar gravity modes
are excited by the Goldstone field source as horizon crossing is 
approached. On superhorizon scales, as $k\to 0,$ the radiation fluid
modes are unexcited. 

The degree of phase dispersion is determined from ratio of the eigenvalues
of the matrix $F=\pmatrix{ F_{jj}&F_{jy}\cr  F_{yj}&F_{yy}\cr }.$ For
$(k\eta )\ltorder 1,$ where our calculation is reliable, $F$ is roughly
proportional to $\pmatrix{1.0&0.17\cr 0.17&0.05\cr }.$ Because of the 
dominance of the $F_{jj}$ matrix element, the eigenvectors very nearly 
coincide with the `growing'
and `decaying' modes ${j_1(k\eta /\sqrt{3})/ \eta }$ and
${y_1(k\eta /\sqrt{3})/ \eta },$ respectively,
with the `growing' eigenvalue dominating over the `decaying'
eigenvalue by a factor of approximately $49.$ Thus the amplitude
of the other mode is approximately $15\% $ that of the dominant
mode, which almost coincides with the growing mode with very little
mixing.    

\vskip -.5in
\def\Onne{1}
\LoadFigure\Onne{\baselineskip 13 pt
\noindent\narrower\ninerm  
The plotted curves (from top to bottom) represent the
matrix elements $F_{jj},$ $F_{jy},$ and $F_{yy}$ at $\eta =1$ as a function
of wavenumber $k.$}
{\epsfysize 3truein}{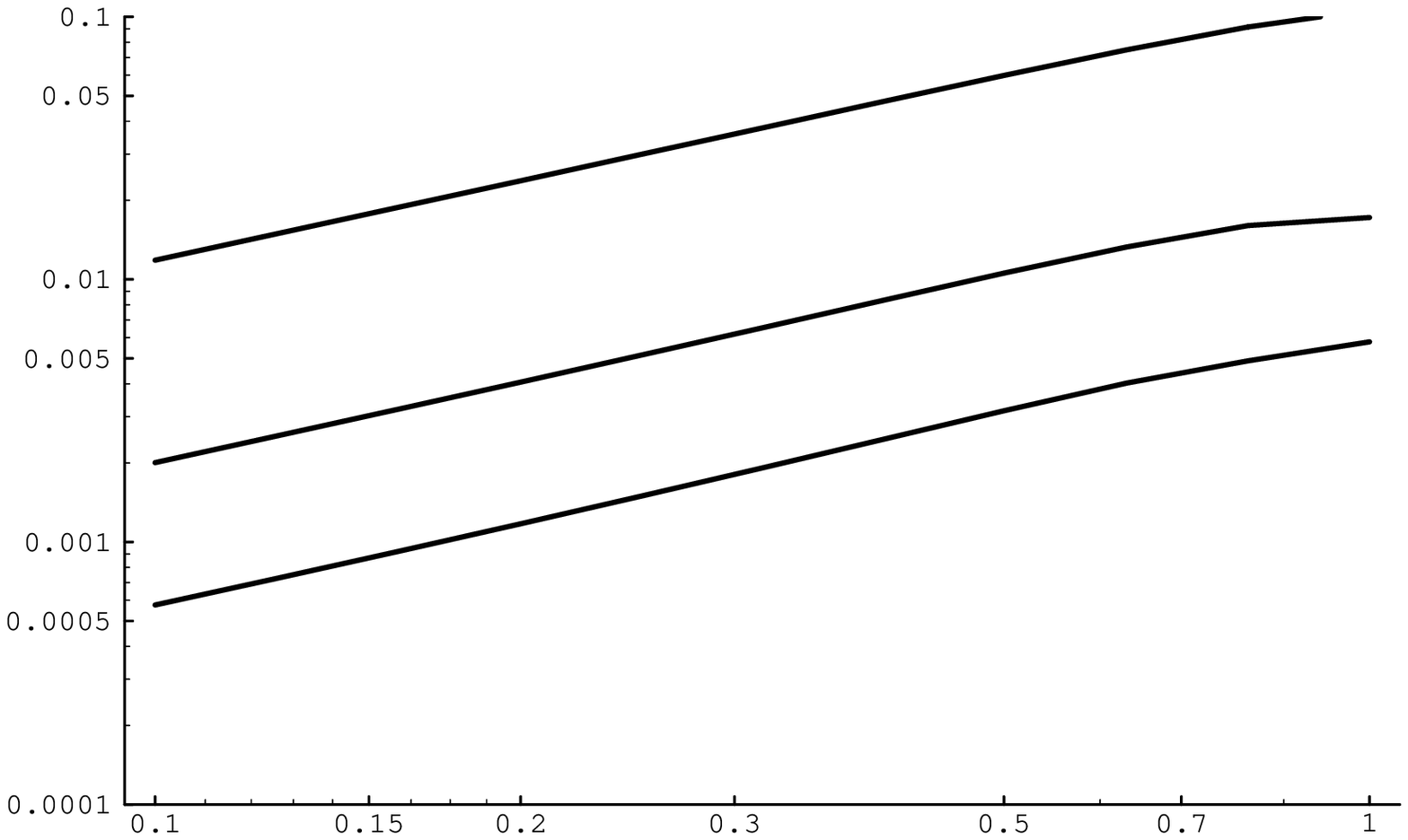}{}

\chapter{Discussion}

We have presented a class of plausible inflationary models with
Goldstone modes that generate non-Gaussian density perturbations
that are comparatively easy to compute. It is of great interest
to determine whether the primordial density perturbations of our 
universe inferred from observation are Gaussian, as predicted by the 
simplest inflationary models, or whether they are of a more
complicated, non-Gaussian variety. Since Gaussian models comprise
an infinitesimal fraction of all possible cosmological models,
in order to define meaningful tests of Gaussianity to consider
such tests in the abstract is insufficient. Instead proposed
tests of Gaussianity must be judged on their ability to rule
out plausible non-Gaussian cosmological models. 

In all non-Gaussian models based on symmetry-breaking phase
transitions---the so-called topological defect models---nonlinearity 
plays a key role, making the predictions of such models 
difficult to compute. By contrast, as we have shown, in many
cases inflationary Goldstone modes may be linearized with 
very little error, making the computation of $N$-point density 
perturbations a matter of evaluating Feynman diagrams.

One shortcoming of the calculations in this paper,
one which we plan to remedy in future work, is that we have
retained only the Goldstone modes corresponding to growing 
modes on superhorizon scales. Because of the neglect of 
the decaying modes of the underlying Goldstone field, the calculations
are reliable only on superhorizon scales and cannot be trusted
on subhorizon scales. To extend the calculation to include the
decaying modes should pose no real difficulty, and with this
extension it should be possible to compute all sorts of 
correlation functions by evaluating Feynman diagrams. One
should be able to compute the CMB moments in this way and
higher-order correlation functions, such as the three-point
function and beyond. Another future direction involves making 
sky maps of the CMB anisotropy, a problem for which 
computing N-point functions would not be practical and for 
which some sort of nonperturbative renormalization in position space would
be required.  
 
Within the limitations due to neglecting the decaying
modes, we were able to determine that, somewhat surprisingly, 
there is very little phase dispersion. The absence of
significant phase dispersion suggests that the CMB moments
from the inflationary linearized Goldstone mode theory should 
exhibit a series of well-defined Doppler peaks. 

After this work was completed, we became aware of an interesting
related paper \refmark{\mukhanov } that discusses non-Gaussian 
perturbations from inflation generated by a massive scalar field. 

{\bf Acknowledgements:} We would like to thank 
A. Linde, G. Sterman, N. Turok, A. Vilenkin, and F. Wilczek
for useful discussions. This work was supported in part by 
the David and Lucile Packard Foundation and by 
National Science Foundation grant PHY 9309888. 

\refout 

\APPENDIX{A}{A---Green's Functions for a Stiff Source}

We work in Newtonian gauge (which is equivalent to the gauge 
invariant formalism)\refmark{\cpma }
and for the stress-energy in addition to 
the stiff source described by the divergenceless tensor 
$\Theta _{\mu \nu }$ assume a single-component perfect
fluid with sound speed $c_s.$ The line element is $ds^2=a^2(\eta )\cdot 
[\eta _{\mu \nu } +h_{\mu \nu }]dx^\mu ~dx^\nu $ and $\eta _{\mu \nu }
={\rm diag}[-1,+1,+1,+1].$

For {\it scalar} perturbations, $h_{00}=-2\phi ,$ 
$h_{0i}=0,$ and $h_{ij}=-2\psi ~
\delta _{ij},$ and the linearized scalar Einstein equations read as follows:
$$\eqalignno{
\delta G^0_0&={-2\over a^2}\cdot [ \nabla ^2\psi -3\H (\dot \psi +\H \phi )]\cr
&=(8\pi G)\cdot [-\rho _c~\delta +\Theta ^0_0],&(A1a)\cr 
\delta G^0_i&={-2\over a^2}\cdot [ \dot \psi +\H \phi ]_{\vert i}\cr
&=(8\pi G)\cdot [\rho _c(1+w)v_{\vert i}+\Theta ^0_i], &(A1b)\cr 
\delta G^j_i(L)&={2\over a^2}\cdot [ 
\ddot \psi +2\H \dot \psi +\H \dot \phi +(2\dot \H +\H ^2)\phi +
{1\over 3}\nabla ^2(\phi -\psi )]~\delta ^j_i\cr
&=(8\pi G)\cdot [\rho _cc_s^2~\delta ~\delta ^j_i+\Theta ^j_i(L)],&(A1c)\cr 
\delta G^j_i(T)&={-1\over a^2}\cdot [ 
(\nabla _i\nabla ^j-{1\over 3}\delta _i^j~\nabla ^2)(\phi -\psi )]~
\cr
&=(8\pi G)\cdot \Theta ^j_i(T)&(A1d)\cr }$$
where (L) and (T) denote the longitudinal and transverse {\it scalar} parts of
the spatial-spatial tensors, respectively. Here the 
dots indicate derivatives with respect to conformal time, $\H =(\dot a /a),$
$\rho _c=(3/8\pi G)\H ^2 a^{-2},$ and $v_i=v_{\vert i}$ 
(i.e., $v$ is the potential for 
the {\it scalar} part of the velocity field).

By adding eqn. (A1a) multiplied by $c_s^2$ to eqn. (A1c),
one obtains
$$\eqalign{ 
&\ddot \psi +(2\H +3\H c_s^2)\dot \psi +\H \dot \phi 
+(2\dot \H +\H ^2+3\H^2c_s^2)\phi 
+{1\over 3}\nabla ^2(\phi -\psi )-c_s^2\nabla ^2\psi \cr
&~~~~~~= (4\pi G)\cdot [\Theta _L-c_s^2\Theta _{00}] .\cr }
\eqno{(A2)}$$
Here $\Theta _{ij}(L)=\delta _{ij}\Theta _L.$ Working in momentum space
and using eqn. (A1d) to eliminate $\phi $ in favor of $\psi ,$ we may write 
$$\eqalign{
&\ddot \psi (\k )+3\H (1+c_s^2)~\dot \psi (\k )+(2\dot \H +\H ^2+3\H^2c_s^2)~
\psi (\k )
+c_s^2k^2 \psi (\k )\cr
&~~~~~~= (4\pi G)\cdot \biggl[ \Theta _L(\k )-c_s^2\Theta _{00}(\k )\cr 
&~~~~~~~~~~~~~~~~~~~~~~~
+{2\over 3}\Theta _T(\k )-{2\H \over k^2}~\dot \Theta _T(\k )-
{4\dot \H+2\H ^2(1+3c_s^2)\over k^2}~\Theta _T(\k )
\biggr] .\cr }
\eqno{(A3)}$$
Here we have adopted the convention 
$$\Theta _{ij}(T)(\k )=\left( 
{k_ik_j-{1\over 3}k^2\delta _{ij}\over k^2}\right) 
\Theta _T(\k ).\eqno{(A4)}$$

For the special case of a completely radiation-dominated universe,
so that $a(\eta )=\eta $ and $c_s^2={1\over 3},$
eqn. (A3) becomes 
$$\eqalign{
&\ddot \psi (\k )+{4\over \eta }\dot \psi (\k )
+{k^2\over 3}\psi (\k )\cr 
&~~~~=(4\pi G)\cdot \left[ \Theta _L(\k )-{1\over 3}\Theta _{00}(\k ) 
+{2\over 3}\Theta _T(\k )-{2\over k^2\eta }\dot \Theta _T(\k )
\right] .\cr }
\eqno{(A5)}$$
which is essentially a Bessel equation of order $\nu ={3\over 2}.$
The homogeneous version of this equation has the general solution
$\psi (\k ,\eta )=\eta ^{-1}\bigl[ \psi (\k ,j)~j_1(k\eta /\sqrt{3})
+\psi (\k ,y)~y_1(k\eta /\sqrt{3})\bigr] .$ From this homogeneous
solution we may construct Green functions, and eqn. (A5)
is solved by 
$$\psi (\k ,\eta )=\eta ^{-1}\bigl[ 
\psi (\k ,\eta ,j)j_1 (k\eta /\sqrt{3})
+\psi (\k ,\eta ,y)y_1(k\eta /\sqrt{3})\bigr] 
\eqno{(A6)}$$
where $j_1(x)={\sin x/x^2}-{\cos x/x},$ 
$y_1(x)=-{\cos x/x^2}-{\sin x/x},$
$$\eqalign{
\psi (\k &,\eta ,j)={-k\over \sqrt{3} }\int _0^\eta d\bar \eta ~
\bar \eta ^{3}~y_1(k\bar \eta /\sqrt{3})\cr
& \times (4\pi G)\cdot \left[ \Theta _L(\k ,\bar \eta )-{1\over 3}
\Theta _{00}(\k ,\bar \eta )
+{2\over 3}\Theta _T(\k ,\bar \eta )-{2\over k^2\bar \eta }
\dot \Theta _T(\k ,\bar \eta )
\right] \cr }
\eqno{(A7a)}$$
and 
$$\eqalign{
\psi (&\k ,\eta ,y)={+k\over \sqrt{3} }\int _0^\eta d\bar \eta ~
\bar \eta ^{3}~j_1(k\bar \eta /\sqrt{3})\cr
& \times (4\pi G)\cdot \left[ \Theta _L(\k ,\bar \eta )-{1\over 3}
\Theta _{00}(\k ,\bar \eta )
+{2\over 3}\Theta _T(\k ,\bar \eta )-{2\over k^2\bar \eta }
\dot \Theta _T(\k ,\bar \eta )
\right] .\cr }
\eqno{(A7b)}$$

\end